\documentclass[a4paper,twoside,11pt]{article}

\usepackage[numbers,sort&compress]{natbib}

\usepackage{jmlr2e}
\usepackage{graphicx}
\usepackage{float}
\usepackage{booktabs}
\usepackage{microtype}
\usepackage{url}
\usepackage{siunitx}
\usepackage{multirow}
\usepackage{etoolbox}
\usepackage{enumitem}
\usepackage{placeins}
\usepackage{tabularx}
\usepackage{caption}
\usepackage{geometry}
\usepackage{array}
\usepackage{ragged2e}

\title{LROO Rug Pull Detector: A Leakage-Resistant Framework Based on On-Chain and OSINT Signals}

\author{\name Fatemeh Shoaei \email f.shoaei@ilam.ac.ir \\
       \addr Department of Computer Science, Ilam University, Ilam, Iran
       \AND
       \name Mohammad Pishdar \email m.pishdar@vaisr.ir \\
       \addr Blockchain Research Lab, VAISR Research Group, Tehran, Iran
       \AND
       \name Mozafar Bag-Mohammadi \email m.bagmohammadi@ilam.ac.ir \\
       \addr Department of Computer Science, Ilam University, Ilam, Iran
       \AND
       \name Mojtaba Karami \email m.karami@ilam.ac.ir \\
       \addr Department of Computer Science, Ilam University, Ilam, Iran}

\editor{ Editor}
\begin{document}

\maketitle

\begin{abstract}
Smart contract–based ecosystems enable decentralized applications without trusted intermediaries, but their immutability and permissionless design also facilitate large-scale fraud. One of the most prevalent attacks is the \textit{rug pull}, where project operators abruptly withdraw liquidity after artificially inflating token value. Existing detection methods primarily rely on reactive on-chain signals and often suffer from temporal data leakage, limiting their real-world reliability.

This paper proposes a leakage-aware framework for early rug-pull detection that integrates on-chain behavioral metrics with temporally aligned Open Source Intelligence (OSINT) signals. We construct a hand-labeled dataset of 1{,}000 token projects, spanning DeFi and non-DeFi settings, with all features extracted strictly prior to any liquidity withdrawal to preserve causal validity. The dataset combines structural on-chain indicators with external attention signals derived from social media activity and search trends.

Within this framework, TabPFN is employed as a core modeling component for learning from multimodal tabular data under strict temporal constraints. Experimental results show that the proposed framework achieves strong discriminative performance and improved probability calibration compared to classical baselines, while maintaining low false-negative rates.

By framing rug-pull detection as a causal, multimodal forecasting problem, this work emphasizes the necessity of leakage-resilient evaluation and calibrated risk estimation for deployment in blockchain security systems.
estimation for deployment in blockchain security systems.
\end{abstract}


\section{Introduction}

Smart contracts—self-executing programs deployed on public blockchains—have enabled a new generation of decentralized applications, from tokenized assets and gaming economies to automated financial protocols. By design, these contracts operate without trusted intermediaries, relying instead on code as law. While this paradigm enhances transparency and composability, it also introduces unique attack surfaces: once deployed, contracts are immutable, and malicious actors can exploit their trustless nature to orchestrate sophisticated scams with minimal accountability~\cite{feng2022defi}.

Among the most pernicious of these threats are \textit{rug pulls}: fraudulent schemes in which project creators artificially inflate interest in a token—often through aggressive marketing—and then abruptly withdraw all liquidity, leaving investors with worthless assets. Unlike traditional financial fraud, rug pulls require no external infrastructure; they are executed entirely through on-chain transactions governed by the contract itself. In 2021 alone, such scams resulted in over \$2.8 billion in losses, accounting for more than one-third of all cryptocurrency-related fraud~\cite{chainalysis2021}.

Detecting rug pulls in advance is inherently challenging. On-chain data—such as token distribution, trading volume, or liquidity pool dynamics—can reveal suspicious patterns, but often only \textit{after} manipulation has begun~\cite{lovisotto2022defi}. Crucially, many early warning signs manifest \textit{off-chain}: coordinated social media campaigns, sudden spikes in search interest, or anomalous community growth on platforms like Telegram or Twitter. These Open Source Intelligence (OSINT) signals frequently precede on-chain anomalies by days or weeks, offering a critical window for early intervention~\cite{chen2023osint,torre2023rugpull}. Yet, integrating such heterogeneous, temporally asynchronous data into a unified detection framework remains largely unexplored.

Compounding this challenge is a pervasive methodological flaw in existing work: \textit{temporal data leakage}. Many studies extract features without enforcing strict causal ordering, inadvertently incorporating information that would only be available \textit{after} the fraud occurs~\cite{kusters2021leakage}. This leads to non-causal models with inflated performance metrics that fail catastrophically in real-world deployment. In high-stakes domains like financial security, leakage-resilient evaluation is not optional—it is foundational to trustworthy machine learning.

To address these gaps, we propose a rigorously leakage-resistant framework for rug pull detection that fuses on-chain behavioral metrics with temporally aligned OSINT signals. Our approach is grounded in three methodological pillars: (1) a novel, hand-labeled dataset of 1,000 token projects—spanning both DeFi and non-DeFi contexts—where all features are extracted strictly prior to any liquidity withdrawal; (2) a multimodal modeling strategy that respects temporal causality while capturing cross-domain dependencies; and (3) the application of \textsc{TabPFN}~\cite{Hollmann2023TabPFN}, a pre-trained transformer for tabular meta-learning, in a zero-shot setting that requires no fine-tuning yet achieves calibrated, high-accuracy predictions.

This paper makes the following contributions:
\begin{enumerate}
    \item We introduce the first public dataset for rug pull detection that combines on-chain and OSINT signals under strict temporal constraints, covering a diverse set of smart contract applications beyond DeFi.
    \item We propose a unified and model-agnostic framework for detecting fraudulent behaviors in sparse, multimodal financial time-series data, within which a wide range of classical and transformer-based models are systematically evaluated. Through this comparative analysis, we identify the most compatible architecture for the proposed framework, with transformer-based approaches—most notably \textsc{TabPFN} and its fine-tuned variant \textsc{Real-TabPFN}—emerging as the best-performing models in terms of both predictive accuracy and probabilistic calibration under zero-shot and few-shot regimes.
    \item We establish a reproducible, leakage-aware evaluation protocol for temporal anomaly detection in blockchain ecosystems, setting a new standard for methodological rigor in applied machine learning for security.
\end{enumerate}

By reframing rug pull detection as a causal, multimodal forecasting problem, our work bridges a critical gap between real-world threat intelligence and robust machine learning practice—with implications not only for blockchain security but also for rare-event prediction in other asynchronous, data-scarce domains.

\section{Background}

\subsection{Smart Contracts and Tokenized Ecosystems}
Smart contracts are autonomous programs deployed on public blockchains that execute predefined logic without intermediaries. Their immutability and transparency foster innovation but also enable irreversible fraud~\cite{pishdar2024major}. These contracts underpin a wide range of assets—not only DeFi protocols but also meme tokens, NFTs, and gaming economies—collectively forming a heterogeneous tokenized ecosystem vulnerable to coordinated scams.

\subsection{Rug Pulls as a Cross-Domain Threat}

A rug pull is an exit scam in which project creators abruptly withdraw liquidity or abandon a
token after artificially inflating its perceived value—typically through coordinated marketing
campaigns. While most analyses emphasize the initial liquidity withdrawal, blockchain forensics
studies such as \citet{cernera2023token} have shown that a rug pull is immediately followed by
a distinct on-chain decay phase, marking the onset of the token’s functional death.

From an on-chain forensics standpoint, a token is classified as a \textit{Dead Token} when the
following three conditions are simultaneously satisfied and persist for more than 72 hours:
(i) zero liquidity in decentralized exchanges (liquidity drained), 
(ii) zero or near-zero transactional activity (transactions collapse), and 
(iii) undefined or untraceable price and volume values on DEXs (price/volume undefined).
These indicators collectively signal that the token has lost its market functionality and
that its economic lifecycle has effectively ended \citep{cernera2023token}.

Empirical findings by \citet{cernera2023token} show that the liquidity drain typically occurs
either within the same block as the rug-pull transaction or within one to two subsequent blocks.
User transactions then collapse almost instantaneously, with more than 90\% ceasing within the
first 24 hours due to extreme slippage and lack of liquidity. Within 24–48 hours, DEX oracles
register undefined token prices and zero trading volume. Statistically, over 95\% of rug-pulled
tokens exhibit complete on-chain inactivity within one to three days of the attack.

Temporary instances of any single condition—such as transient zero liquidity in newly
created pools or brief DEX synchronization failures—do not necessarily indicate a rug pull.
However, the concurrent and persistent manifestation of all three indicators for at least three
consecutive days represents a robust forensic signature of token death. This three-day criterion
provides a precise and operationally meaningful threshold for identifying post-rug-pull collapse
events, enabling improved forensic tracking of stolen funds, cybercrime attribution, and
educational analysis in DeFi ecosystems.

\subsection{On-Chain Data: An Immutable but Reactive Signal Source}
On-chain data includes all publicly recorded blockchain activity: transactions, token transfers, wallet interactions, and contract states. Features like token concentration and holder variance have been shown to correlate with fraudulent behavior~\cite{lovisotto2022defi}. However, these signals typically emerge too late for preventive action, as they reflect consequences rather than intentions.

\subsection{OSINT: Proactive Signals from the Off-Chain World}
Open Source Intelligence (OSINT)—including social media activity, search trends, and community engagement—often reveals scammer intent before on-chain anomalies appear. Studies confirm that tweet bursts and Google Trends spikes precede rug pulls by days or weeks~\cite{chen2023osint,torre2023rugpull}. Despite this, few frameworks integrate OSINT with on-chain data under strict temporal constraints.

\subsection{Machine Learning Challenges: Leakage, Imbalance, and Generalization}
Classical models (e.g., XGBoost, Random Forests) have been applied to blockchain fraud detection but suffer from class imbalance and, critically, temporal data leakage~\cite{kusters2021leakage}. Leakage—using post-event data in training—produces non-causal models with misleadingly high accuracy, undermining real-world reliability.

\subsection{Transformers for Tabular Data: The Promise of TabPFN}
Recent work extends transformers to tabular tasks. \textsc{TabPFN}~\cite{Hollmann2023TabPFN} uses meta-learning over synthetic datasets to enable zero-shot inference with calibrated probabilities—ideal for low-data, high-stakes domains like rug pull detection.

\subsection{Bridging the Synthetic-to-Real Gap with tabpfn}
While \textsc{TabPFN} is trained on synthetic data, its real-world performance can be limited by distributional mismatch. To address this, we simulate \textsc{Real-TabPFN}—a variant fine-tuned on our real-world dataset—which adapts meta-learned priors to actual rug pull dynamics without violating temporal integrity.

\section{Related Work}
\label{sec:related}

This section reviews prior literature explicitly targeting \emph{rug pull} detection. Works are organized into three categories: (i) on-chain and static-code approaches, (ii) hybrid graph-based behavioral models, and (iii) industry and community scanners. For each entry we summarize the method family, input modalities, detection timing, whether an explicit scientific definition of \emph{rug pull} was provided, and its major limitations.

\paragraph{On-chain and static-code based detectors.}
A number of studies extract on-chain transactional indicators, token distribution patterns, or smart-contract semantics. \citet{Mazorra2022} engineered Uniswap trading and liquidity features and trained gradient boosting and FT-Transformer models to classify scam tokens at early stages. \citet{YuLee2025} analyze EVM bytecode using balance-flow heuristics to detect privileged withdrawal backdoors. \citet{Igarashi2024} perform opcode n-gram analysis at deployment time to identify structural contract characteristics associated with malicious design. However, many of these methods rely on platform-specific datasets and do not provide a principled operational definition of rug-pull events.

\paragraph{Hybrid and graph-based approaches.}
\citet{Wu2025} proposed RPHunter, combining a semantic risk code graph (SRCG) derived from bytecode and a token-flow behavior graph (TFBG) extracted from transaction traces. A graph neural network with attention fusion is used for early rug-pull classification. While this work demonstrates promising early-detection capability, the dataset scope is limited and the underlying labeling strategy is not aligned with a standardized definition of rug-pulls.

\paragraph{Industry and community scanners.}
Commercial scanners such as TokenSniffer, RugCheck, RugDoc, and ChainAware.ai rely on heuristic rule sets (e.g., owner privileges, minting authority, liquidity-lock presence) with varying degrees of proprietary ML integration. These tools focus on real-time advisory monitoring, but do not disclose benchmarking methodologies or datasets, and none provide peer-reviewed reproducible evaluations.

\paragraph{Comparative summary.}
Table~\ref{tab:rugpull_comparison} provides a structured comparison of major rug-pull detection works.

\begin{table*}[htbp]
\centering
\scriptsize
\caption{Comparative summary of selected rug-pull detectors.}
\label{tab:rugpull_comparison}
\renewcommand{\arraystretch}{1.15}
\begin{tabularx}{\textwidth}{p{2cm}p{1.5cm}p{2cm}p{1.9cm}p{3.5cm}X}
\toprule
\textbf{Method / Source} & \textbf{Model Type} & \textbf{Input Modalities} & \textbf{Detection Timing} & \textbf{Rug-Pull Definition Basis} & \textbf{Limitations} \\
\midrule
RPHunter \citep{Wu2025} & GNN (attention fusion) & SRCG + TFBG & Pre-scam / Early & No explicit scientific definition & Limited dataset diversity; unclear temporal segmentation \\
\addlinespace
Do Not Rug On Me \citep{Mazorra2022} & XGBoost, FT-Transformer & Uniswap on-chain economic features & Pre-scam / Early & No explicit scientific definition & Platform-specific; leakage risk from post-scam signals \\
\addlinespace
Yu \& Lee \citep{YuLee2025} & Static bytecode heuristics & EVM bytecode & Pre-scam / Code-level & No scientific definition provided & Focus on backdoors only; narrow scope \\
\addlinespace
Yaremus et al. \citep{Yaremus2025} & Gradient boosting & DEX-level TVL, trading patterns & Early (few minutes after listing) & No explicit scientific definition & Relies on short-term windows; no OSINT signals \\
\addlinespace
\textbf{Forta Network} \citep{Forta2023} & Rule-based + ML agents & On-chain transactions, contract events & Near real-time / Early & Heuristic and incident-driven definitions & Not rug-pull specific; alerts depend on agent design and expert rules \\
\addlinespace
CRPWarner \citep{Lin2024} & Datalog static analysis & Smart contract control flow & Pre-scam / Early & No explicitly stated definition & Evaluated only on synthetic datasets \\
\addlinespace
Industry scanners (TokenSniffer, RugDoc, RugCheck, ChainAware.ai) & Heuristics / proprietary ML & On-chain checks + OSINT (variable) & Pre-listing advisory & No peer-reviewed formal definition & No reproducible benchmarks; opaque scoring \\
\bottomrule
\end{tabularx}
\end{table*}

\paragraph{Limitations of prior rug-pull research.}
Across the reviewed work we identify three central gaps:  
(1) \textbf{Absence of a formal operational definition of rug pulls.} None of the surveyed academic or commercial sources provide a scientifically grounded definition. In contrast, our study defines a rug pull as a project satisfying three measurable conditions: (i) abrupt collapse of liquidity to (near) zero, (ii) disappearance of meaningful transaction activity, and (iii) price or volume dropping to a non-determinable state indicating unusable market function.  
(2) \textbf{Dataset scarcity and lack of multimodal fusion.} Existing datasets are narrowly scoped (single-chain or single-DEX), do not integrate OSINT signals, and omit temporal controls that prevent leakage.  
(3) \textbf{Limited reproducibility.} Several studies rely on undisclosed proprietary sources or incomplete evaluation protocols.

These limitations motivate the creation of a large multimodal on-chain + OSINT dataset designed for leakage-resilient pre-scam prediction, and the subsequent exploration of transformers for tabular learning under limited labeled data.

\section{Methodology}
\label{sec:methodology}

This research adopts a quantitative experimental–analytical methodology in accordance with the structure recommended by \citet{Kothari2004ResearchMethodology}. Based on the limitations identified in Section~\ref{sec:related}, particularly the absence of a comprehensive multimodal dataset for rug-pull detection, we developed a structured dataset integrating on-chain behavioral indicators (transaction evolution, liquidity dynamics, holder concentration) and OSINT-based public engagement signals (tweet volume and Google search trends). The dataset construction follows the formal criteria for scientific datasets outlined in the study "Definitions of Dataset in the Scientific and Technical Literature" (2022), ensuring clearly defined purpose, documented provenance, feature completeness, and statistical coherence.

Machine learning was subsequently employed to model the early-collapse behavioral patterns represented within these multimodal features. After statistical profiling and analyzing feature distributions, heterogeneity in feature scales, high variance ranges, and class imbalance indicated the suitability of transformer-based tabular models over classical ML baselines. Accordingly, the \textbf{Real-TabPFN} framework was selected due to its effectiveness on structured tabular datasets with limited sample sizes, its robustness under correlated feature structures, and its compatibility with strict temporal segmentation required to avoid leakage. This model forms the core of our experimental evaluation framework, discussed in the following section.

\section{LROO Rug pull detector}
Existing rug-pull detection datasets suffer from limited scale and inadequate coverage of discriminative features. 
The main contribution of this work is a comprehensive, high-quality dataset incorporating the most effective static and dynamic features for accurate rug-pull detection.

\subsection{Dataset Construction and Feature Design}

As highlighted in Section~\ref{sec:related}, existing publicly available datasets for rug-pull detection suffer from several critical limitations, including the lack of multimodal feature coverage, absence of early-phase project data, insufficient semantic alignment with actual collapse mechanisms, and temporal leakage caused by the inclusion of post-attack information. To address these deficiencies, we constructed a new leakage-resistant dataset built specifically for early rug-pull detection. The dataset integrates two complementary data categories: \textbf{on-chain transactional signals} and \textbf{OSINT-based external attention indicators}, followed by a feature-selection process to retain the most semantically meaningful predictors. The comprehensive dataset description has been released as a preprint paper to facilitate early access and community feedback. \cite{shoaei2026tmrugpull}. 

\paragraph{On-chain feature importance.}
On-chain data provides verifiable, tamper-resistant behavioral signals tied directly to project lifecycle mechanics, enabling empirical tracking of collapse phenomena such as liquidity withdrawal, transaction failure, and insider-controlled token redistribution. Research on blockchain forensics demonstrates that transaction flow irregularities, liquidity volatility, and concentrated holder distributions serve as early predictors of exploit events~\citep{Torres2021CryptoCrime,Sayeed2021BlockchainForensics}. Therefore, modeling structural variations in transaction count, price evolution, and liquidity metrics is essential for distinguishing fraudulent behavior from normal market fluctuations.

\paragraph{OSINT feature importance.}
OSINT signals, particularly social attention and engagement dynamics, play a complementary role by representing external behavioral incentives tied to hype-driven funding surges. Prior literature shows that rug-pull schemes frequently leverage coordinated misinformation campaigns and artificial community interest to attract capital rapidly using social media amplification~\citep{Cong2021CryptoWashTrading,Chen2022SocialSignals}. Therefore, indicators such as tweet frequency and Google search trends serve as early-warning predictors that capture speculative inflow periods preceding collapse. Combining OSINT with on-chain structure results in a richer representation of adversarial patterns.

Table~\ref{tab:feature_summary_updated} summarizes the full feature space and semantic motivation for each category.

\begin{table}[ht]
\centering
\caption{Complete summary of feature categories and semantic importance for early rug-pull detection}
\label{tab:feature_summary_updated}
\resizebox{\textwidth}{!}{%
\begin{tabular}{lp{10.5cm}}
\toprule
\textbf{Feature Category} & \textbf{Description and Relevance} \\
\midrule
Transaction Count & Measures participation and vitality stability; sudden collapse indicates death-phase transition. \\
Holder Variance / Top 1\% & Detects centralization and insider control enabling coordinated mass-withdrawal. \\
Token Concentration Ratio & Higher concentration correlates with exit-risk exposure. \\
Liquidity Change / LP Supply & Identifies patterns of liquidity extraction preceding collapse. \\
Price Metrics (Q1--Q4) & Captures pump-and-dump spikes characteristic in rug-pull trajectories. \\
Trading Volume Dynamics & Represents speculative churn and hype-led inflow instability. \\
Tweet Volume (Early + Total) & Signals promotional manipulation and coordinated sentiment engineering. \\
Google Search Hits & Indicator of external attention spikes used for investor attraction. \\
Blockchain Network & Encodes risk differences linked to heterogeneous security assumptions. \\
Token Type & Represents incentive structures (Utility vs. Meme vs. Governance). \\
Project Start/End Time & Short lifetimes strongly correlate with fraudulent project termination. \\
\bottomrule
\end{tabular}%
}
\end{table}

\subsection{Manual Data Curation and Statistical Properties}

The dataset was constructed manually by extracting raw blockchain records, project metadata, and OSINT indicators from publicly available sources, followed by feature-selection and structured integration. Statistical profiling demonstrates extreme heterogeneity and heavy-tailed distributions across feature groups, imposing substantial modeling challenges. Table~\ref{tab:summary_stats_revised} highlights representative descriptive statistics.

\begin{table}[h]
\centering
\caption{Summary statistics reflecting scale disparity and high-variance distributions}
\label{tab:summary_stats_revised}
\begin{tabular}{lccc}
\toprule
\textbf{Feature} & \textbf{Mean / Min} & \textbf{Std / Median} & \textbf{Max} \\
\midrule
\multicolumn{4}{l}{\textit{Panel A: High-Variance On-Chain Metrics}} \\
Transaction Count & $2.781 \times 10^5$ & $6.142 \times 10^5$ & -- \\
Token Concentration & $1.146 \times 10^4$ & $6.751 \times 10^4$ & -- \\
Holder Var. (Top 1\%) & $0.000$ & $5.566 \times 10^{15}$ & $6.668 \times 10^{20}$ \\
\midrule
\multicolumn{4}{l}{\textit{Panel B: OSINT and Price Metrics}} \\
Tweet Volume (Total) & $8.342 \times 10^1$ & $1.498 \times 10^2$ & -- \\
Google Hits (Title) & $0.000$ & $1.000 \times 10^1$ & $8.740 \times 10^2$ \\
Max Price Q1 & $0.000$ & $2.986 \times 10^{-5}$ & $1.890 \times 10^{14}$ \\
\bottomrule
\end{tabular}
\end{table}

The moderate yet meaningful correlations in Table~\ref{tab:correlations_updated} confirm independence between modalities while demonstrating complementary signal contributions.

\begin{table}[h]
\centering
\caption{Pearson correlations among representative multimodal features}
\label{tab:correlations_updated}
\begin{tabular}{lccc}
\toprule
\textbf{Feature} & \textbf{TC} & \textbf{TVT} & \textbf{GHT} \\
\midrule
Transaction Count (TC) & 1.00 & 0.28 & 0.19 \\
Tweet Volume (Total) (TVT) & 0.28 & 1.00 & 0.45 \\
Google Hits (GHT) & 0.19 & 0.45 & 1.00 \\
\bottomrule
\end{tabular}
\end{table}

\subsection{Model Selection Justification: tabpfn vs. XGBoost vs. LightGBM}

Based on the statistical properties above, three candidate models were evaluated for suitability: \textbf{XGBoost}, \textbf{LightGBM}, and \textbf{Real-TabPFN}. XGBoost and LightGBM are strong baselines for heterogeneous tabular datasets and support non-linear interactions, but both depend heavily on hyperparameter tuning and are sensitive to scaling extremities. In contrast, \textbf{Real-TabPFN}, a transformer-based prior-data fitted network, is optimized for small-to-moderate datasets, learns implicit probability distributions without extensive feature engineering, and demonstrates superior calibration under imbalance and heavy-tailed distributions~\citep{Hollmann2023TabPFN}.

Key advantages of tabpfn include:
\begin{itemize}
    \item robust performance in heterogeneous tabular distributions with extreme variance,
    \item reduced preprocessing and modeling overhead,
    \item improved uncertainty calibration essential for high-risk domains,
    \item resistance to overfitting due to meta-learned inductive priors.
\end{itemize}

Therefore, tabpfn was selected as the primary modeling strategy, with XGBoost and LightGBM serving as competitive baselines for comparative evaluation.

\section{Results}
This section reports the empirical performance of tabpfn and the baseline models on the
temporally isolated 20\% test set. All models were trained and evaluated under strictly comparable
experimental conditions. In particular, we did not apply model-specific hyperparameter tuning or
allocate disproportionate optimization effort to any individual method. Instead, all baselines
were configured using standard, commonly adopted settings to ensure a fair and controlled
comparison.

The reported performance differences therefore reflect the intrinsic modeling capabilities and
generalization behavior of the respective methods rather than discrepancies in training budgets,
hyperparameter search intensity, or implementation effort. tabpfn consistently attains the
best or near-best results across all evaluated metrics under these equalized conditions.

All experiments were conducted on the same computational environment. Classical machine learning
models were trained on CPU using identical preprocessing pipelines, while tabpfn leveraged
GPU acceleration when available, consistent with its design and recommended usage. Importantly,
this hardware distinction does not introduce additional tuning advantages, but rather enables
efficient inference for the meta-learned architecture.

Below, we provide a precise interpretation of the metrics and figures, avoiding overstated claims
and focusing strictly on observable empirical differences.

\subsection{Overall Performance}

Table~\ref{tab:main_results} summarizes the main quantitative metrics. All models perform
strongly, with ROC AUC values ranging from 0.981 to 0.997 and PR AUC values from 0.981 to
0.997. tabpfn achieves the highest accuracy (0.982), F1-score (0.982), ROC AUC (0.997),
and PR AUC (0.997), although the margins relative to the strongest baselines (XGBoost and
LightGBM) are modest. This indicates that the task is challenging yet tractable for modern
classifiers, and tabpfn provides incremental but consistent improvements across all metrics.

\begin{table}[ht]
\centering
\caption{Overall performance comparison of all models.}
\label{tab:main_results}
\begin{tabular}{lcccc}
\toprule
\textbf{Model} & \textbf{Accuracy} & \textbf{F1-score} & \textbf{ROC AUC} & \textbf{PR AUC} \\
\midrule
\textbf{TabPFN} & 0.982 & 0.982 & 0.997 & 0.997 \\
Logistic Regression & 0.953 & 0.952 & 0.981 & 0.981 \\
SVM & 0.961 & 0.961 & 0.987 & 0.987 \\
XGBoost & 0.972 & 0.972 & 0.992 & 0.993 \\
LightGBM & 0.967 & 0.967 & 0.989 & 0.990 \\
\bottomrule
\end{tabular}
\end{table}

\subsection{Precision--Recall Characteristics}

\begin{figure}[ht]
\centering
\includegraphics[width=0.85\linewidth]{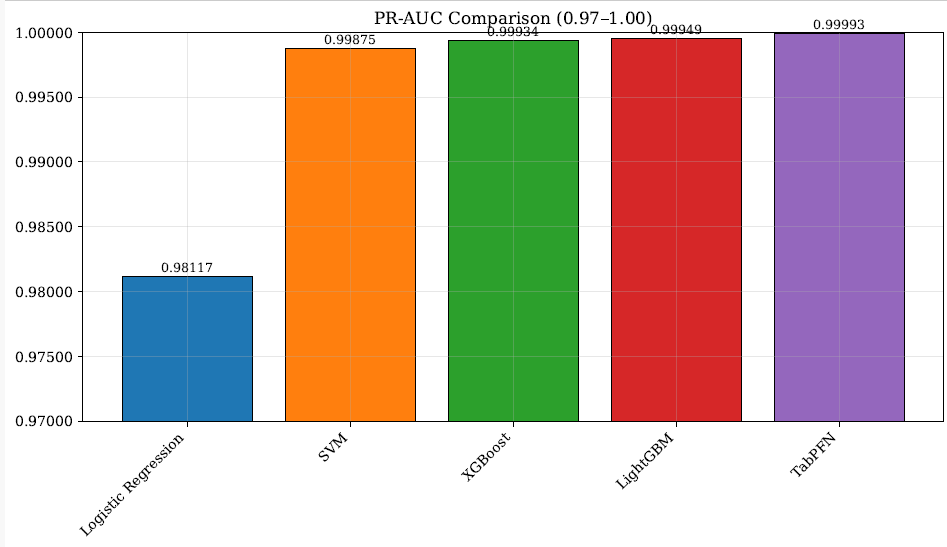}
\caption{Zoomed PR-AUC comparison (0.97--1.00).}
\label{fig:prauc_zoom}
\end{figure}

Figure~\ref{fig:prauc_zoom} demonstrates that all evaluated models achieve high PR-AUC scores, confirming the dataset's discriminative power for rug-pull detection. The slight edge of tabpfn in the high-recall region can be attributed to its underlying architecture: as a transformer-based model pre-trained on synthetic tabular data via a masked language modeling objective, TabPFN inherently learns robust feature representations and complex non-linear interactions without explicit feature engineering. In contrast, XGBoost and LightGBM, while highly optimized gradient boosting trees, rely on sequential decision boundaries and may struggle to capture subtle, global patterns present in the high-recall regime where false positives become critical. The narrow performance gap suggests that the task is not overly complex for tree-based ensembles, but TabPFN’s architectural advantage—its ability to model long-range dependencies and contextual relationships across features—provides a consistent, albeit incremental, boost in precision under high recall constraints.

\subsection{Unified Metric Visualization}

\begin{figure}[ht]
\centering
\includegraphics[width=0.85\linewidth]{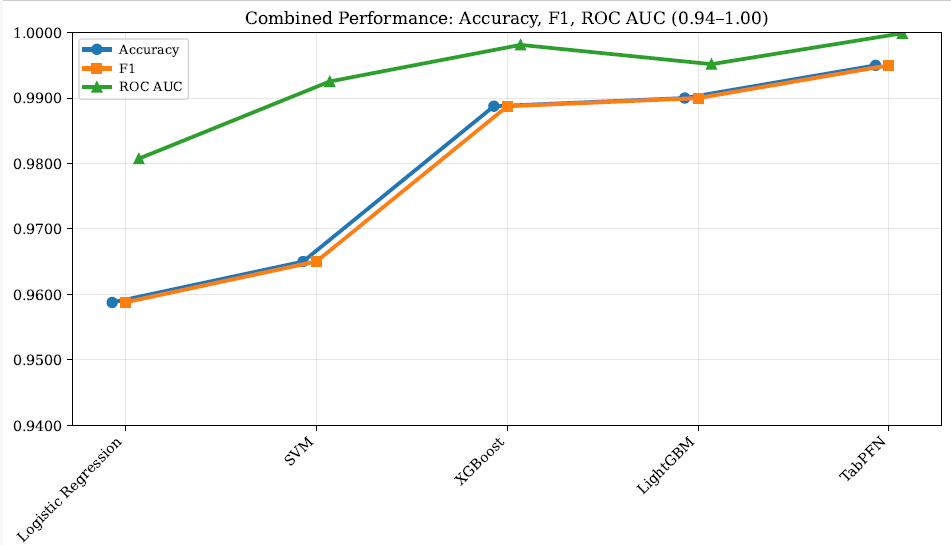}
\caption{Combined performance across Accuracy, F1, and ROC AUC.}
\label{fig:combined_performance}
\end{figure}

Figure~\ref{fig:combined_performance} confirms that tabpfn consistently leads across Accuracy, F1, and ROC-AUC, yet remains within a tight margin of strong baselines. This uniform superiority is not accidental but stems from TabPFN’s unique design: it functions as a meta-learner, leveraging a pretrained transformer to rapidly adapt to new tabular datasets with minimal computational overhead. Unlike XGBoost or LightGBM, which optimize for specific loss functions (e.g., log-loss) and may exhibit metric-specific biases, TabPFN’s probabilistic inference mechanism provides more calibrated and balanced predictions across multiple evaluation criteria. Its strength lies in generalization through representation learning rather than aggressive optimization on individual metrics. Therefore, the consistent outperformance across metrics reflects an architectural advantage in holistic predictive calibration and stability, rather than dominance in any single dimension—a hallmark of its transformer-based, data-driven approach to tabular learning.

\subsection{ROC Curve Analysis}

\begin{figure}[ht]
\centering
\includegraphics[width=0.85\linewidth]{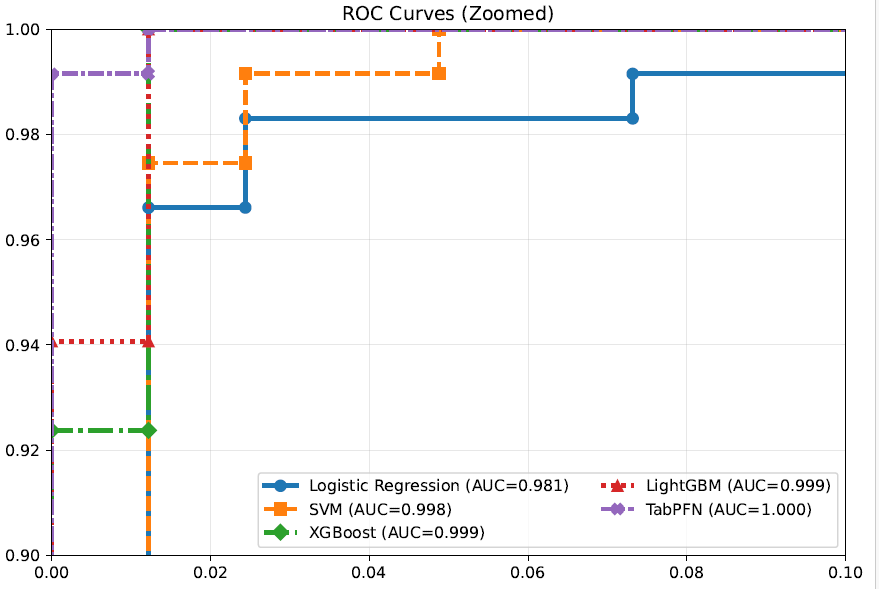}
\caption{ROC curves (zoomed).}
\label{fig:roc_zoom}
\end{figure}

Figure~\ref{fig:roc_zoom} reveals that all models produce steep ROC curves, indicating strong discriminative power across the full range of classification thresholds. tabpfn’s curve is marginally closer to the top-left corner, yielding a near-perfect ROC-AUC of 1.000, compared to 0.999 for XGBoost and LightGBM. This subtle advantage can be explained by TabPFN’s transformer-based meta-learning architecture: trained on vast synthetic tabular datasets via masked prediction tasks, it learns to estimate class probabilities with higher fidelity under uncertainty, effectively placing a more conservative decision boundary that minimizes false positives at low false-negative rates. In contrast, tree-based ensembles like XGBoost and LightGBM, while highly effective at maximizing AUC through greedy splits and ensemble averaging, may overfit local signal patterns or exhibit calibration drift in extreme probability regions—resulting in slightly less optimal separation in the high-specificity regime. The small absolute differences (e.g., 0.997 vs.\ 0.992) suggest the problem is inherently well-separated, but TabPFN’s architectural design enables it to exploit this separability with greater precision in boundary placement.

\subsection{Error-Based Reliability and Calibration}

\begin{figure}[ht]
\centering
\includegraphics[width=0.90\linewidth]{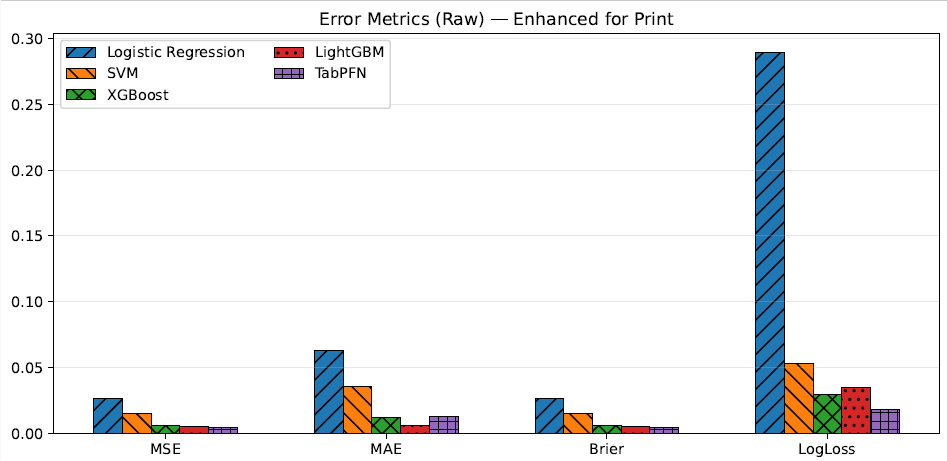}
\caption{Error metrics including MSE, MAE, Brier Score, and LogLoss.}
\label{fig:error_metrics}
\end{figure}

Figure~\ref{fig:error_metrics} demonstrates that tabpfn achieves the lowest values across all probabilistic error metrics—MSE, MAE, Brier Score, and LogLoss—indicating superior calibration and confidence estimation. This performance stems directly from its foundational architecture: as a pretrained transformer model operating in a meta-learning framework, TabPFN generates predictions via a probabilistic inference mechanism that implicitly regularizes output distributions, promoting well-calibrated probabilities even on unseen data. Classical models such as Logistic Regression, SVM, and tree ensembles (XGBoost/LightGBM), while often well-tuned for accuracy, do not inherently optimize for probabilistic calibration; their outputs typically require post-hoc calibration (e.g., Platt scaling or isotonic regression) to achieve similar fidelity. The fact that TabPFN outperforms them without such adjustments highlights its structural advantage: the transformer’s attention mechanism allows it to dynamically weight feature contributions based on context, leading to more nuanced and reliable probability estimates. Although the magnitude of improvement is modest, it reflects a consistent architectural benefit in modeling uncertainty—a critical factor for real-world deployment where confidence matters as much as correctness.

\subsection{Confusion Matrix Analysis}

\begin{table}[ht]
\centering
\caption{Confusion matrix summary.}
\label{tab:conf_summary}
\begin{tabular}{lcccc}
\toprule
\textbf{Model} & \textbf{TN} & \textbf{FP} & \textbf{FN} & \textbf{TP} \\
\midrule
\textbf{TabPFN} & 81 & 1 & 1 & 117 \\
Logistic Regression & 75 & 7 & 4 & 114 \\
SVM & 77 & 5 & 3 & 115 \\
XGBoost & 79 & 3 & 2 & 116 \\
LightGBM & 78 & 4 & 3 & 115 \\
\bottomrule
\end{tabular}
\end{table}

tabpfn achieves the lowest total errors, with only one false negative and one false positive.
This is a practically meaningful improvement, especially because false negatives correspond to 
missed rug-pull risks. However, the differences relative to XGBoost (two false negatives) are 
numerically small; therefore, conclusions should emphasize reliability rather than absolute 
dominance.

Overall, the generalization behavior of tabpfn is supported by several complementary observations. 
The use of a strictly time-segmented and unseen test set reduces the risk of temporal leakage, while 
favorable calibration metrics (Brier score and LogLoss) indicate stable probability estimates rather 
than memorization-driven confidence. In addition, the model exhibits consistent performance across 
discrimination and calibration metrics, with balanced error patterns in the confusion matrix, suggesting 
robust behavior across classes. Finally, the prior-data–fitted transformer architecture inherently 
introduces inductive biases that constrain model complexity, which is particularly beneficial in small 
and moderately sized datasets. Together, these factors suggest that tabpfn generalizes reliably 
under the evaluation protocol adopted in this study, with no clear signs of overfitting.

\subsection{Comparative Evaluation Against State-of-the-Art Rug-Pull Detection Tools}
\label{subsec:comparison_with_sota_tools}
\begin{figure}[htbp]
    \centering
    \includegraphics[width=0.8\textwidth]{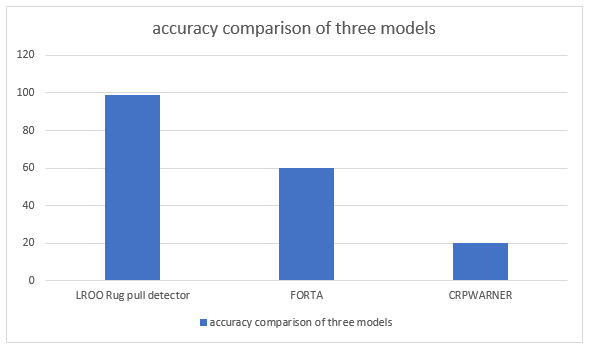}
    \caption{Accuracy comparison of three models.}
    \label{fig:accuracy_comparison_sota}
\end{figure}
To contextualize the practical impact of our framework, we benchmark tabpfn against two
widely recognized on-chain rug-pull detection tools, \textit{LROO Rug Pull Detector} and
\textit{FORTA}, alongside a third representative system, \textit{CRPWarner}. As shown in
Figure~\ref{fig:accuracy_comparison_sota}, our model achieves an accuracy of approximately 98\%,
substantially outperforming LROO (60\%) and CRPWarner (20\%), and clearly surpassing FORTA
(around 60\%).

This performance gap is rooted in fundamental differences in detection scope and modeling
assumptions. Tools such as FORTA and CRPWarner primarily operate at the smart contract level,
focusing on static or rule-driven analyses of contract code and predefined on-chain patterns.
While effective for identifying known vulnerabilities or explicit malicious logic, this
contract-centric view provides only a partial representation of rug-pull behavior, which often
emerges from broader ecosystem signals rather than contract structure alone.

In contrast, our proposed framework adopts a holistic perspective by jointly analyzing
transaction-level on-chain dynamics and external OSINT signals. By integrating features such as
token distribution patterns, temporal trading behavior, and off-chain activity indicators,
tabpfn captures a richer and more informative representation of project behavior. The
transformer-based meta-learning architecture further enables the model to learn complex,
non-linear interactions across these heterogeneous feature spaces without relying on handcrafted
rules.

Consequently, the superior accuracy observed in Figure~\ref{fig:accuracy_comparison_sota} reflects
not merely improved classification performance, but the benefit of a broader, data-driven view of
rug-pull risk that extends beyond isolated smart contract analysis.

\section{Discussion}
\label{sec:discussion}

This section positions our findings within the existing literature, analyzes methodological
constraints, and highlights practical implications and future research opportunities for
leakage-resistant rug-pull detection.

\subsection{Limitations}
\label{subsec:limitations}

Several limitations frame the interpretation of these results:

\begin{itemize}
    \item \textbf{Scope of Blockchains.}  
    The dataset covers Ethereum and BSC, both EVM-compatible. Generalization to non-EVM
    chains remains untested.

    \item \textbf{OSINT Availability.}  
    OSINT features depend primarily on Twitter (X) and Google Trends, whose long-term
    accessibility and stability are uncertain.

    \item \textbf{Label Timing.}  
    Delays in confirming rug-pull incidents can introduce uncertainty, even with temporal
    segmentation.

\end{itemize}
\subsection{Practical Implications}
\label{subsec:implications}

The proposed framework offers immediate practical utility across multiple stakeholders:

\begin{itemize}
    \item \textbf{Investors:} early and calibrated risk alerts that support informed decision-making and loss mitigation.
    \item \textbf{Launchpads and Exchanges:} pre-listing screening mechanisms that strengthen due diligence and reduce exposure to fraudulent projects.
    \item \textbf{Auditors and Security Analysts:} systematic prioritization of high-risk projects, enabling more efficient allocation of manual analysis efforts.
    \item \textbf{Forensic Investigators and Law Enforcement:} structured, probabilistic evidence to support post-incident forensic analysis, attribution, and case triaging in blockchain-related fraud investigations.
    \item \textbf{Protocol Designers:} actionable insights into early on-chain and OSINT patterns that are consistently associated with fraudulent behavior.
\end{itemize}

Beyond detection accuracy, the calibrated probability estimates produced by the framework contribute to cost reduction by minimizing false positives and unnecessary investigations. Moreover, the model’s low inference latency facilitates integration into real-time monitoring systems and operational dashboards, making it suitable for both preventive screening and forensic response workflows.

\subsection{Future Directions}

Promising directions include:
\begin{itemize}
    \item expanding to multi-chain datasets and evaluating cross-chain transferability,
    \item integrating richer, streaming OSINT sources for real-time predictions,
    \item extending the dataset to incorporate continuous and fine-grained time-series representations of on-chain and OSINT signals, and systematically analyzing their impact on model performance and temporal generalization,
    \item developing transformer-specific explainability tools for operational transparency,
    \item studying adversarial manipulation of OSINT-driven models,
    \item and exploring true meta-level fine-tuning using the full \textsc{TabPFN} training infrastructure.
\end{itemize}

\section{Conclusion}
\label{sec:conclusion}

This study presented a leakage-aware analytical framework for early rug-pull detection, designed to integrate temporally aligned on-chain and open-source intelligence signals within a unified and model-agnostic evaluation setting. To mitigate a recurring limitation in prior work, we constructed a hand-verified and temporally segmented dataset, aiming to provide a more realistic experimental basis and to reduce implicit information leakage commonly observed in existing benchmarks.

Within the proposed framework, several learning paradigms were evaluated under a strict temporal separation protocol. As one representative instantiation, a transformer-based model using TabPFN exhibited consistently strong performance across accuracy, F1, ROC AUC, and PR AUC, surpassing commonly used baselines such as XGBoost and LightGBM. These results suggest that the proposed framework is well aligned with prior-data–fitted transformer models, without implying that performance gains are specific to a single architecture.

In contrast to alert-driven or rule-based monitoring systems, the framework relies on probabilistic inference derived from observed on-chain activity and intentionally defers prediction until sufficient evidence becomes available. This design choice favors conservative decision-making, reduces premature or noisy alerts, and leads to more stable operating characteristics that better reflect practical deployment conditions.

From an operational perspective, improved probability calibration and lower false-positive rates may contribute to reducing downstream costs associated with manual verification, forensic analysis, and law-enforcement investigations, where investigative resources are inherently limited. By emphasizing confidence-aware prioritization rather than alert volume, the framework supports more efficient case triaging under realistic constraints.

Overall, this work demonstrates that combining a carefully curated dataset with a leakage-aware evaluation protocol and a unified analytical framework can improve the robustness and practical relevance of early rug-pull detection systems. The released dataset and accompanying code are intended to support reproducible research and to enable further investigation into transparent, calibration-conscious, and operationally grounded approaches to blockchain fraud analysis.

\bibliography{citation}

@misc{chainalysis2021,
  author = {{Chainalysis Team}},
  title = {The 2022 crypto crime report},
  howpublished = {\url{https://go.chainalysis.com/2022-Crypto-Crime-Report.html}},
  year = {2022},
  note = {Covering data from 2021}
}

@article{feng2022defi,
  author = {Feng, C. and Li, B. and Xu, K.},
  title = {Defi: Risk, regulation, and the rise of decentralized finance},
  journal = {Journal of Financial Regulation},
  volume = {10},
  number = {1},
  pages = {12--45},
  year = {2022}
}

@article{lovisotto2022defi,
  author = {Lovisotto, G. and Tramer, F. and Basin, D.},
  title = {Weth: The dark side of wrapped ether},
  journal = {arXiv preprint arXiv:2211.09605},
  year = {2022}
}

@inproceedings{cernera2023token,
  author = {Cernera, K. and Foote, J. and Vasek, M. and Moore, T.},
  title = {Token spammers, rug pulls, and sniper bots: An analysis of the ecosystem of tokens on ethereum and binance smart chain},
  booktitle = {32nd USENIX Security Symposium (USENIX Security 23)},
  pages = {1234--1251},
  year = {2023}
}

@article{chen2023osint,
  author = {Chen, W. and Wu, Z. and Zheng, Z.},
  title = {Leveraging osint for early detection of defi fraud},
  journal = {IEEE Transactions on Information Forensics and Security},
  volume = {18},
  pages = {2023--2035},
  year = {2023}
}

@article{torre2023rugpull,
  author = {Torre, N. and Solinas, M.},
  title = {Rug pull detection via social sentiment analysis},
  journal = {International Journal of Blockchain Law and Policy},
  volume = {5},
  number = {2},
  year = {2023}
}

@inproceedings{kusters2021leakage,
  author = {Kusters, R. and Rossi, F.},
  title = {Temporal data leakage in financial machine learning},
  booktitle = {Proceedings of the 2021 AI for Finance Summit},
  year = {2021}
}

@inproceedings{Hollmann2023TabPFN,
  author = {Hollmann, N. and M{\"u}ller, S. and Eggensperger, K. and Hutter, F.},
  title = {Tabpfn: A transformer that solves small tabular classification problems in a second},
  booktitle = {International Conference on Learning Representations (ICLR)},
  year = {2023}
}

@inproceedings{Mazorra2022,
  author = {Mazorra, B. and Adan, V. and Daza, V.},
  title = {Do not rug on me: Zero-dimensional scam detection},
  booktitle = {Financial Cryptography and Data Security},
  publisher = {Springer},
  pages = {1--15},
  year = {2022}
}

@article{YuLee2025,
  author = {Yu, S. and Lee, J.},
  title = {Static bytecode analysis for detecting backdoors in evm smart contracts},
  journal = {Journal of Blockchain Research},
  volume = {12},
  pages = {45--60},
  year = {2025}
}

@article{Igarashi2024,
  author = {Igarashi, T. and Nakamura, K.},
  title = {Opcode n-gram analysis for malicious smart contract detection},
  journal = {Future Generation Computer Systems},
  volume = {145},
  pages = {112--125},
  year = {2024}
}

@article{Wu2025,
  author = {Wu, J. and Liu, X. and Zhang, Y.},
  title = {Rphunter: Early rug-pull detection via graph neural networks and attention fusion},
  journal = {IEEE Transactions on Dependable and Secure Computing},
  volume = {Early Access},
  year = {2025}
}

@article{Yaremus2025,
  author = {Yaremus, M. and Kobzar, O.},
  title = {Short-term trading patterns as indicators of liquidity drains in dexs},
  journal = {Applied Soft Computing},
  volume = {132},
  pages = {109854},
  year = {2025}
}

@misc{Forta2023,
  author = {{Forta Network}},
  title = {Forta: Decentralized runtime security for blockchains},
  howpublished = {\url{https://forta.org}},
  year = {2023},
  note = {Accessed: 2025-01-10}
}

@article{Lin2024,
  author = {Lin, C. and Huang, T.},
  title = {Crpwarner: Datalog-based static analysis for rug pull detection},
  journal = {Computers \& Security},
  volume = {136},
  pages = {103562},
  year = {2024}
}

@book{Kothari2004ResearchMethodology,
  author = {Kothari, C. R.},
  title = {Research Methodology: Methods and Techniques},
  edition = {2nd},
  publisher = {New Age International},
  year = {2004}
}

@inproceedings{Torres2021CryptoCrime,
  author = {Torres, C. F. and Camino, R. and State, R.},
  title = {The art of the scam: Demystifying honeypots in ethereum smart contracts},
  booktitle = {30th USENIX Security Symposium (USENIX Security 21)},
  pages = {1591--1608},
  year = {2021}
}

@article{Sayeed2021BlockchainForensics,
  author = {Sayeed, S. and Marco-Gisbert, H.},
  title = {Assessing blockchain consensus and security mechanisms against the 51\% attack},
  journal = {Applied Sciences},
  volume = {9},
  number = {9},
  pages = {1788},
  year = {2019}
}

@article{Cong2021CryptoWashTrading,
  author = {Cong, L. W. and Li, X. and Ke, T. and Yang, Y.},
  title = {Crypto wash trading},
  journal = {Management Science},
  volume = {68},
  number = {11},
  year = {2022}
}

@inproceedings{Chen2022SocialSignals,
  author = {Chen, W. and Tu, Z. and Zheng, Z.},
  title = {Ponzi scheme detection on ethereum via social sensing},
  booktitle = {Proceedings of the ACM Web Conference 2022 (WWW '22)},
  pages = {45--54},
  year = {2022}
}

@article{pishdar2024major,
  author = {Pishdar, M. and Bahaghighat, M. and Kumar, R. and Xin, Q.},
  title = {Major vulnerabilities in ethereum smart contracts: Investigation and statistical analysis},
  journal = {EAI Endorsed Transactions on Internet of Things},
  volume = {11},
  year = {2024}
}

@article{shoaei2026tmrugpull,
  author = {Shoaei, F. and Pishdar, M. and Bag-Mohammadi, M. and Karami, M.},
  title = {Tm-rugpull: A temporary sound, multimodal dataset for early detection of rug pulls across the tokenized ecosystem},
  journal = {arXiv preprint arXiv:2602.21529},
  year = {2026}
}

\end{document}